\documentclass[preprint, eqsecnum, aps]{revtex4} 
\usepackage{amsmath}
\usepackage{amsfonts}
\newcommand{\nwc}{\newcommand}
\nwc{\be}{\begin{equation}}
\nwc{\ee}{\end{equation}}
\nwc{\bea}{\begin{eqnarray}}
\nwc{\eea}{\end{eqnarray}}
\nwc{\mR}{{\mathbb{R}}}
\nwc{\mS}{{\mathbb{S}}}
\nwc{\mZ}{{\mathbb{Z}}}
\nwc{\mC}{{\mathbb{C}}}
\nwc{\mI}{{\mathbb{I}}}
\nwc{\cH}{{\cal H}}
\nwc{\cB}{{\cal B}}
\nwc{\cR}{{\cal R}}
\nwc{\cU}{{\cal U}}
\nwc{\Om}{\Omega}
\nwc{\cC}{{\cal C}}
\nwc{\cD}{{\cal D}}
\nwc{\rP}{{\rm P}}
\nwc{\chii}{{\underline {\overline \chi}}}

\begin{document}
\title{Wigner distributions for finite dimensional quantum systems: 
An algebraic approach}
\author{S Chaturvedi\dag\footnote{To whom correspondence should be addressed 
(scsp@uohyd.ernet.in)}, E Ercolessi\ddag, G Marmo\S, 
G Morandi$\|$, \mbox{N Mukunda}\P\ and R Simon$^+$}

\address{\dag School of Physics, University of Hyderabad, Hyderabad 500 046, 
India}

\address{\ddag Physics Department, University of Bologna, INFM and INFN, Via 
Irnerio 46, I-40126, Bologna, Italy}

\address{\S Dipartimento di Scienze Fisiche, University of Napoli and INFN, 
Via Cinzia, I-80126 Napoli, Italy}

\address{$\|$ Physics Department, University of Bologna, INFM and INFN, V.le 
B.Pichat 6/2, I-40127, Bologna, Italy}

\address{\P Centre for High Energy Physics, Indian Institute of Science, 
Bangalore 560 012, India}

\address{$^+$ The Institute of Mathematical Sciences, C.I.T. Campus, Chennai 
600113, India}

\begin{abstract}
We discuss questions pertaining to the definition of `momentum', `momentum
space', `phase space', and `Wigner distributions'; for finite dimensional 
quantum systems. For such systems, where traditional concepts of `momenta' 
established for continuum situations offer little help, we propose a 
physically reasonable and mathematically tangible definition and use it for 
the purpose of setting up Wigner distributions in a purely algebraic manner. 
It is found that the point of view adopted here is limited to odd dimensional 
systems only. The mathematical reasons which force this situation 
are examined in detail.
\end{abstract}
\maketitle
\section{Introduction}
The description of states of quantum systems with the help of Wigner
distributions, in the case of continuous Cartesian position and 
momentum variables \cite{1}-\cite{3}, has found applications in a wide 
variety of contexts. These include the original aim of calculating quantum 
corrections to classical statistical mechanics, various aspects of quantum 
optics, and quantum chemistry \cite{4}. 

There has been for some time considerable interest in developing a similar
method for quantum systems possessing a finite dimensional state space 
\cite{5}-\cite{12}. In the works of Feynman\cite{6} and of Wootters\cite{5}, 
which constitute some of the early efforts in this direction, it was found 
necessary to treat independently the cases of two-dimentional state space, 
and of an odd prime dimensional state space. In the former, corresponding 
to the case of a qubit, the insights from the treatment of a spin 1/2 
angular momentum in quantum mechanics were used 
effectively. The treatment of the odd prime dimension case has been geometric 
in spirit. From the start it is assumed that the phase space is a 
$\underline{{\rm square}}$  array of points, namely that `momentum space' is 
of the same `size' as `position' or configuration space. In this phase space, 
the notions of straight lines, parallel lines, and foliations or striations 
 are set up and used in stating the properties desired of a Wigner 
distribution and then a solution is proposed \cite{5},\cite{11}. The 
arithmetic of a finite field of odd prime order is used in the constructions. 
For general odd dimensions or a power of two, one has to take the Cartesian 
product of the elementary solutions. 

The purpose of the present work is to provide an alternative treatment of this
problem based on an algebraic approach. As we will find, apart from
reproducing known results (in the odd dimension cases), essentially new
possibilities will also come to light. Our work is motivated by the following
points of view. In elementary mechanics, the primitive notion of momentum is
as a numerical measure of rate of change of position in space, the factor of
proportionality being the mass. In canonical mechanics we move to an algebraic
role for momentum, as generator of translations in configuration space, with
mass not playing an explicit role. In Cartesian quantum mechanics this
algebraic role is expressed by the canonical Heisenberg commutation relations
and is fundamental. 

Returning to classical canonical mechanics usually the configuration space $Q$
is taken to be a differentiable manifold of some dimension. Then the concepts
of momenta and phase space are given immediately by the automatic construction
of the cotangent bundle $T^\star Q$. There are of course special features in 
the Cartesian case $Q=\mR^n$, not shared by the general case. Then 
$T^\star Q=\mR^{2n}$, and both
positions and conjugate momenta are elements of real linear vector spaces of
the common dimension, $n$. Thus they are of the same algebraic nature,
allowing us to form real linear combinations of them in a physically
significant manner. This leads to an important role for the groups $Sp(2n,R)$
in the classical case, and to the double covers $Mp(2n)$ after
quantisation. For a general configuration space $Q$ different from $\mR^n$, for
instance $\mS^n$ or the manifold of a Lie group\cite{13}, one sees that 
coordinates and
momenta have intrinsically different algebraic natures. Locally momenta always
belong to a linear vector space, while positions may have no such
characterisation. In these cases, of course, quantum mechanics uses suitably
square integrable functions on $Q$ as wave functions, and the basic operator
relations are inspired by what one knows about the classical $T^\star Q$. 
However, already in the simple case of $Q=\mS^2$, a particle confined to 
the surface of the unit sphere, one sees an interesting new possibility 
emerging. Since $T^\star \mS^2$ is
not parallelisable, the canonical momenta cannot be globally defined as
variables. However one can view $\mS^2$ as the coset space 
$SO(3)/SO(2)$, and so think of canonical momenta as the constrained components 
of a three dimensional angular momentum, obeying the algebraic requirement of 
being orthogonal to the (unit) position vector. This makes the momenta 
globally well defined, besides bringing into the picture a role for the 
group $SO(3)$ acting transitively on $\mS^2$. 

We now turn to an $N$-level quantum system. The state space is a complex
$N$-dimensional Hilbert space, wave functions are complex $N$-component column
vectors, but properly speaking no notion of canonical momenta is given in
advance in any intrinsic sense. There are no fundamental operator relations
analogous to the canonical Heisenberg commutation relations. In this case,
classically we may view the configuration space $Q$ as a finite set of $N$
discrete points, without any concept of canonical momenta. This leads to the
viewpoint: for finite level systems, the problems of defining momenta,
momentum space and phase space should be regarded as  part of the problem of
setting up Wigner distributions in a physically reasonable manner. 

From this discussion we extract and explore the idea that for any choice of
configuration space $Q$, whether a differentiable manifold or a finite
discrete set, the concept of momentum is $\underline{{\rm any~ group}}$~$G$~$
\underline{{\rm that~
acts~transitively~ on}}$~$Q$. In case $Q$ is a differentiable 
manifold and $G$ acts on $Q$ via diffeomorphisms, we move a step closer to 
the usual meaning of momentum by
consideration of the generators of these diffeomorphisms subject to as many
linear algebraic constraints as the difference in the dimensions of $G$ and
$Q$. 

This general viewpoint means that $Q$ is a coset space $G/H$ with respect to
some subgroup $H\subset G$. We will however assume for definiteness that $H$ is
trivial, i.e., $Q=G$. Thus the configuration space is itself a group. We will
find that using the representation theory of $G$, in the quantum case this
leads to an algebraic way of defining momenta and phase space. However when we
turn to the problem of defining Wigner distributions the method is limited to
odd dimensions only for a reason that will emerge later. 

This work is organised as follows. In Section 2, following the line of thought
outlined above, we take a group  $G$ of order N as the `classical'
configuration space underlying an $N$-level quantum system; and proceed 
to  set up the necessary machinery for introducing the notions of `positon 
space', `momentum space' and `a point in phase space', recapitulating, in the
process, relevant aspects of the representation theory of finite groups. 
A noteworthy feature of the phase space thus obtained is that    
it is always an $N\times N$ array for any $N$ regardless of the nature of $G$.
Having developed the notion of `phase space' for an $N$-level quantum system, 
for any $N$, in Section 3, we take up the question of constructing  a Wigner 
distribution on such a phase space. This requires introducing the concept 
of a `square root' of a group element, which, as it turns out, is possible
only when $N$ is odd. Restricting ourselves to such dimensions, we show that, 
in general, there are two equally good ways of defining Wigner distributions
which coincide only when $G$ is an abelian group. 
Choosing one of them,  and keeping $G$ general, 
we construct the corresponding phase point operators \cite{5},\cite{14} and
find, in particular, that the phase point operator at the origin, like its
 continuum counterpart, has the satisfying feature of being an inversion 
operator. We then discuss in some detail the special case  when $G$ is 
abelian and analyse the two possibilities (a) $N$ an odd prime (b) $N$ an odd
composite. While in the former case, $G$ gets uniquely fixed to be a cyclic 
group of order $N$, in the latter case, in general, one has more than one
choice for $G$ leading to distinct definitions of the Wigner distribution.
(The material presented in this section makes use of the results in 
\cite{12},\cite{13} though with some changes of notation). In Section 4 
we turn to the case when $G$ is nonabelian  and analyse the relation between 
the two choices for the Wigner distribution. We further show that it is
possible to define an `extended' Wigner distribution which contains the two
Wigner distributions as its marginals. However, as is explicitly demonstrated,
this can only be achieved at the  cost of introducing
overcompleteness. Section 5 contains our concluding remarks and further
outlook.      

\section{Finite Order Configuration and Phase Spaces}
Our aim is to describe general states of an $N$-level quantum systems (for odd
$N$) via Wigner distributions. The space of (pure) quantum states is an
$N$-dimensional complex Hilbert space $\cH^{(N)}$, each vector $|\psi>\in 
\cH^{(N)}$ being an $N$-component complex column vector. The corresponding 
or comparison classical
configuration space $Q$ is a finite discrete set of $N$ points. We wish to
regard $Q$ as a finite group $G$ of order $N$. This makes it natural to label
the points of $Q$ as $g,~g^\prime,\cdots,~a,~b,\cdots \in G$, with a 
distinguished `origin' corresponding to the identity
element $e\in G$. In turn in the quantum case we introduce an 
orthonormal basis of `position eigenstates' $|g>$ for $\cH^{(N)}$: 
\bea
&&<g^\prime|g>=\delta_{g^\prime,g},~~g,~g^\prime \in G;\nonumber\\
&& \sum_{g\in G} |g><g|= \mI.
\label{21}
\eea
A general $|\psi>\in \cH^{(N)} $ has a position space wave function
\bea
&&\psi(g)=<g|\psi>,\nonumber\\
&&<\psi|\psi>=||\psi||^2=\sum_{g} |\psi(g)|^2. 
\label{22}
\eea
To move onto the concepts of momentum and phase space, we consider the
transitive action of $G$ on $Q=G$, namely of $G$ on itself. Keeping in mind
the general non-abelian situation, we have two such actions, namely left
translations $L_g$ and right translations $R_g$:
\bea
(L_{g^\prime}\psi)(g)&=&\psi(g^{\prime -1}g),\nonumber\\
L_{g^\prime}|g>&=&|g^\prime~g>;\nonumber\\
(R_{g^\prime}\psi)(g)&=&\psi(g~g^\prime),\nonumber\\
R_{g^\prime}|g>&=&|g~g^{\prime -1}>.
\label{23}
\eea
(In the abelian case, $L_g=R_g$). Both $L_g$ and $R_g$ are unitary, and they
obey 
\bea
L_{g^\prime}~L_{g}&=& L_{g^\prime g},\nonumber\\
R_{g^\prime}~R_{g}&=& R_{g^\prime g},\nonumber\\
L_{g^\prime}~R_{g}&=& R_g~L_{g^\prime}.
\label{24}
\eea
Thus we have two commuting representations-the regular representations of $G$
on $\cH^{(N)}$. We arrive at the notion of `momentum eigen states' by 
performing
their simultaneous complete reduction into irreducibles. For this we recall
well known facts regarding the inequivalent unitary irreducible representations
(UIR's) of $G$. We label them with a symbol $j$, the corresponding dimension
being $N_j$. With respect to some chosen orthonormal basis within each UIR, we
have corresponding unitary representation matrices  
\be
D^{j}(g)= \left(D^{j}_{mn}(g)\right),~~~m,n=1,2,\cdots, N_j.
\label{25}
\ee
The following well known properties and facts will be needed:
\begin{subequations}\label{26}
\bea 
{\rm Composition}~~~~~\sum_{n^\prime} D^{j}_{mn^\prime}(g^\prime)~
D^{j}_{n^\prime n}(g)&=&D^{j}_{mn}(g^\prime g);\label{26a}\\
{\rm Unitarity}~~~~~~~~~\sum_{n} D^{j}_{mn}(g)~D^{j}_{m^\prime n}(g)^*&=&
\delta_{mm^\prime};\label{26b}\\
{\rm Orthogonality}~~\sum_{g} D^{j}_{mn}(g)~
D^{j^\prime}_{m^\prime n^\prime}(g)^*&=&
\frac{N}{N_j}~\delta_{jj^\prime}~\delta_{mm^\prime}~\delta_{nn^\prime};
\label{26c}\\
{\rm Completeness}~\sum_{jmn}N_j~ D^{j}_{mn}(g)~
D^{j}_{m n}(g^\prime)^*
&\equiv& \sum_{jm}N_j~ D^{j}_{mm}(gg^{\prime-1})\nonumber\\
&=&N\delta_{g,g^\prime};\label{26d}\\
\sum_{j}N_{j}^2&=&N.\label{26e}
\eea
\end{subequations}
It is now natural to define an orthonormal basis of `momentum eigenstates'
for $\cH^{(N)}$, complementary to the basis $\{|g>\}$, as vectors 
labelled $|jmn>$:
\bea
|jmn>&=& \sqrt{\frac{N_j}{N}}\sum_g D^{j}_{m n}(g)|g>,\nonumber\\
<g|jmn>& =& \sqrt{\frac{N_j}{N}}~D^{j}_{m n}(g).
\label{27}
\eea
They obey
\bea
<j^\prime m^\prime n^\prime|jmn>&=& \delta_{j^\prime j}~ 
\delta_{m^\prime m}~\delta_{n^\prime n},\nonumber\\
\sum_{jmn}~|jmn<jmn|&=& \mI,
\label{28}
\eea
while the completely reduced actions of $L_g$ and $R_g$ are :
\bea
L_g~|jmn> &=&\sum_{m^\prime}~D^{j}_{m m^\prime}(g^{-1})~|jm^\prime
n>,\nonumber\\
R_g~|jmn> &=&\sum_{n^\prime}~D^{j}_{n^\prime n}(g)~|jmn^\prime>.
\label{29}
\eea
Therefore `momentum' is `conjugate' to `position' in the sense that when $jmn$
and $g$ are given, a complex number $\sqrt{N_j/N}~D^{j}_{m n}(g)$ is 
determined.

A vector $|\psi>\in \cH^{(N)}$ thus has two complementary wave functions: 
$\psi(g)$ of eq.$(\ref{22})$,
and a momentum space wave function 
\be
\psi_{jmn}=<jmn|\psi>.
\label{210}
\ee
If $|\psi>$ is normalised, then it determines two complementary probability
distributions, as
\be
||\psi||^2=\sum_{g}~|\psi(g)|^2=\sum_{jmn}~|\psi_{jmn}|^2=1.
\label{211}
\ee
For purposes of setting up Wigner distributions, we thus regard a `point in
phase space' as a pair $(g;jmn)$. By eq.$(\ref{26e})$ we then find that the 
`size' of
momentum space is precisely $N$, the same as the `size' of $Q$: so the points
$(g;jmn)$ of phase space form a $\underline{\rm square}$ array. We obtain 
this result here not
as a matter of $\underline{\rm definition}$ but as a $\underline{\rm 
consequence}$ of an important result of group representation
theory. In the phase space there is a distinguished `origin' corresponding to
$(e;000)$: here $g=e$, and $j=m=n=0$ denote the trivial or identity 
representation of $G$.
 
The developments of this section so far are valid for all $N$. Further, each
way of regarding $Q$ as a group of order $N$ leads to a corresponding way of
building up momenta and phase space. Thus if $G$ and $G^\prime$ are two
non isomorphic groups of order $N$, they lead to two distinct notions of
momenta and phase space.

\section{Wigner Distributions}
We begin by recalling an important property available only for odd $N$, i.e.,
a property of finite groups of odd order. It is that in such a group each
element $g\in G$ has a unique `square root' element written as $\sqrt{g}$ 
or $g^{1/2}$:
\be
g\in G \Longrightarrow~~\sqrt{g}\in G~~:~~(\sqrt{g})^2=g.
\label{31}
\ee
The uniqueness and other useful properties of this notion are as follows:
\begin{subequations}
\bea
g_1=g_2 &\Longleftrightarrow &~\sqrt{g_1}= \sqrt{g_2},~~ g_{1}^2=g_{2}^2;
\label{32a}\\
(\sqrt{g})^{-1}&=&\sqrt{g^{-1}};\label{32b}\\
g\sqrt{g} &=& \sqrt{g} g;\label{32c}\\
g^{-1}\sqrt{g}&=&\sqrt{g^{-1}};\label{32d}\\
\sqrt{gg^\prime g^{-1}} &=& g\sqrt{g^\prime} g^{-1}.
\label{32e}
\eea
\end{subequations}
Therefore in any sum over $G$ of a function $f(g)$ belonging to a linear space
we have the freedom to write the sum in different ways:
\be
\sum_{g} f(g) = \sum_{g}f(\sqrt{g})= \sum_{g} f(g^2).
\label{33}
\ee
A further consequence is :
\begin{subequations}\label{34}
\bea
a=g^{\prime-1} g, && b=g^\prime g \Longleftrightarrow \nonumber\\
&& g^\prime=\sqrt{ba^{-1}},~~
g=\sqrt{ab^{-1}}b=\sqrt{ba^{-1}}~a;\label{34a}
\\
a=g g^{\prime-1} , && b=g g^\prime  \Longleftrightarrow \nonumber\\
&& g^\prime=\sqrt{a^{-1}b},~~ g=b\sqrt{b^{-1}a} = a\sqrt{a^{-1}b}.
\label{34b}
\eea
\end{subequations}
In both cases, for given $a$ and $b$, $g$ and $g^\prime$ are unique. In the
nonabelian case, these two equations differ, while in the abelian case they
coincide. We will find that these two properties of any group of odd order are
essential to be able to define Wigner distributions with desirable properties
using the present method. 

We have seen in Section 2 that a point in phase
space is a pair $(g;jmn)$. It is then natural to look for a Wigner
distribution in the form $W_{\psi}(g;jmn)$ for given normalised 
$|\psi>\in \cH^{(N)}$ and require that it
have desirable properties with regard to reality, transformations under left
and right group actions, reproduction of the two marginal probability
distributions in eq.$(\ref{211})$, and traciality. Keeping in mind the general 
case of
nonabelian $G$, it turns out that there are two distinct but equally good ways
of setting up Wigner distributions, which reproduce complementary features of
the momentum space probability distribution $|\psi_{jmn}|^2$. We define them in
sequence. 

\vskip0.2cm
\noindent
{\underline {\rm Wigner distribution I}}
\vskip0.2cm

For any given normalised $|\psi>\in \cH^{(N)}$ we define a corresponding 
Wigner distribution by
\be
W_{\psi}(g;jmm^\prime) =\sum_{g^\prime} \psi(g^{\prime-1}g) 
D^{j}_{m m^\prime}(g^{\prime~ 2})\psi(g^\prime g)^*.
\label{35}
\ee
This definition extends immediately to a general state $\widehat{\rho}$ as
\be
W_{\widehat{\rho}}(g;jmm^\prime) =\sum_{g^\prime} <g^{\prime-1}g|{\widehat \rho}|g^\prime g>
D^{j}_{m m^\prime}(g^{\prime~ 2}),
\label{36}
\ee
however for simplicity we shall work mainly with the pure state case. Using
the properties $(\ref{26},\ref{33},\ref{34})$ as necessary, we find : 
\begin{subequations}\label{37}
\bea
W_{\psi}(g;jmm^\prime)^* &=& W_{\psi}(g;jm^\prime m);\label{37a}\\
\psi^\prime = L_{g^\prime} \psi : W_{\psi^\prime}(g;jmm^\prime)&=&
\sum_{m_1,m_{1}^\prime}
D^{j}_{m m_1}(g^{\prime})W_{\psi}(g^{\prime -1}g;jm_1 m_{1}^\prime)\times
\nonumber\\&&~~~~~~~~~~~~~~~~~~~~~~~~~D^{j}_{m_{1}^\prime m^\prime}(g^{\prime-1}),\nonumber\\
\psi^{\prime\prime} = R_{g^\prime} \psi : 
W_{\psi^{\prime\prime}}(g;jmm^\prime)&=&
W_{\psi}(gg^{\prime};jmm^\prime);\label{37b}\\
\frac{1}{N}\sum_g N_j~W_{\psi}(g;jmm^\prime)&=&\sum_n \psi_{jmn}^*
\psi_{jm^\prime n}~,\nonumber\\
\frac{1}{N}\sum_{jm}N_j ~W_{\psi}(g;jmm)&=&|\psi(g)|^2;\label{37c}\\ 
\frac{1}{N}\sum_g\sum_{jmm^\prime} N_j~ W_{\phi}(g;jm^\prime m )
&&W_{\psi}(g;jmm^\prime)=|<\phi|\psi>|^2.
\label{37d}
\eea
\end{subequations}
From the last traciality property it is clear that $|\psi>$ is completely
described by $ W_{\psi}(g;jmm^\prime)$ as far as all quantum mechanical 
probabilities are
concerned. For the marginals, we see that the `position space' probability
distribution $|\psi(g)|^2$ is faithfully reproduced. However, in the 
nonabelian case, the
`momentum space' probability distribution (in a polarised form) is reproduced
in what we may call a `coarse grained' version. This is further discussed in
Section 4. 

Properties $(\ref{37a},\ref{37b})$ and the second of $(\ref{37c})$ are 
evident essentially upon
inspection. The first of eq.$(\ref{37c})$ follows upon use of 
eq.$(\ref{34a})$ and changing
summation variables. For the traciality property $(\ref{37d})$ slightly more 
work is needed. Using eq.$(\ref{26d})$ at the appropriate stage, and then 
eq.$(\ref{34a})$, we get:
\bea
&&\frac{1}{N}\sum_g\sum_{jmm^\prime} N_j W_{\phi}(g;jm^\prime m )
W_{\psi}(g;jmm^\prime)=\nonumber \\
&&\frac{1}{N}\sum_{g,g^\prime,g^{\prime\prime}}\sum_{jmm^\prime}
N_j \phi(g^{\prime-1}g)D^{j}_{m^\prime  m}(g^{\prime~2})
\phi(g^\prime g)^* 
\psi(g^{\prime\prime-1}g)D^{j}_{mm^\prime}(g^{\prime\prime~2})
\psi(g^{\prime\prime} g)^* \nonumber\\
&=& \sum_{g,g^\prime,g^{\prime\prime}} \phi(g^{\prime-1}g)
\psi(g^{\prime\prime} g)^*\psi(g^{\prime\prime-1}g)\phi(g^\prime g)^* 
\delta_{g^{\prime\prime},g^{\prime-1}}
\label{38}
\eea
which immediately gives $(\ref{37d})$.
\vskip0.2cm
\noindent
{\underline {\rm Wigner distribution II}}
\vskip0.2cm
Now in place of eq.$(\ref{35})$ we define a different 
(in the nonabelian case !) Wigner distribution by
\be
W_{\psi}^{~\prime} (g;jnn^\prime) =\sum_{g^\prime} \psi(gg^{\prime-1}) 
D^{j}_{n n^\prime}(g^{\prime~ 2})\psi(g g^\prime)^*.
\label{39}
\ee
Then in place of eqs.$(\ref{37})$ we find :
\begin{subequations}\label{310}
\bea
W_{\psi}^{~\prime}(g;jnn^\prime)^* &=& W_{\psi}^{~\prime}(g;jn^\prime n);
\label{310a}\\
\psi^\prime = L_{g^\prime} \psi : W_{\psi^\prime}^{~\prime}(g;jnn^\prime)&=&
W_{\psi}^{~\prime}(g^{\prime-1}g;jnn^\prime),\nonumber\\
\psi^{\prime\prime} = R_{g^\prime} \psi : 
W_{\psi^{\prime\prime}}^{~\prime}(g;jnn^\prime)&=&
\sum_{n_1,n_{1}^\prime}
D^{j}_{n n_1}(g^{\prime})
W_{\psi}^{~\prime}(gg^{\prime};jn_1
n_{1}^\prime) 
D^{j}_{n_{1}^\prime n^\prime}(g^{\prime-1});\nonumber\\&&\label{310b}\\
\frac{1}{N}\sum_g N_j W_{\psi}^{~\prime}(g;jnn^\prime)&=&\sum_m \psi_{jmn}
\psi_{jmn^\prime}^{*},\nonumber\\
\frac{1}{N}\sum_{jn}N_j W_{\psi}^{~\prime}(g;jnn)&=&|\psi(g)|^2;
\label{310c}\\ 
\frac{1}{N}\sum_g\sum_{jnn^\prime} N_j W_{\phi}^{~\prime}(g;jn^\prime n )
&&W_{\psi}^{~\prime}(g;jnn^\prime)=
|<\phi|\psi>|^2.
\label{310d}
\eea
\end{subequations}

Having exhibited the existence of these two ways of defining the Wigner
distribution distinct for nonabelian $G$, we hereafter mainly work with
definition I, except in Section 4 where we explore the relationships between
the two. 
\vskip0.2cm
\noindent
{\underline {\rm Phase Point Operators}}
\vskip0.2cm
In analogy with earlier treatments, it is possible to express the Wigner
distribution $(\ref{35})$ as the expectation value of a 
`phase point operator' in the state $|\psi>$:
\bea
W_{\psi}(g;jmm^\prime)&=& <\psi|{\widehat W}(g;jmm^\prime)|\psi>,\nonumber\\
{\widehat W}(g;jmm^\prime)&=& \sum_{g^\prime}|g^\prime g>
D_{mm^\prime}^j(g^{\prime~2}) <g^{\prime-1} g|.
\label{311}
\eea
Since with the help of eq.$(\ref{26d})$ and $(\ref{32a})$ we are able to 
`invert' this to read 
\be
|g^{\prime}g><g^{\prime -1}g|=\sum_{jmm^\prime}\frac{N_j}{N} 
~D_{mm^\prime}^j(g^{\prime~2})^*{\widehat W}(g;jmm^\prime),
\label{312}
\ee
we see that they provide a basis for the space of all operators 
on $\cH^{(N)}$. They have the specific properties:
\begin{subequations}
\bea
{\widehat W}(g;jmm^\prime)^\dagger&=& {\widehat W}(g;jm^\prime m);\label{313a}
\\
\frac{N_j}{N}\sum_{g}{\widehat W}(g;jmm^\prime)& =&\sum_{n}|jmn><jm^\prime n|;
\label{313b}\\  
\sum_{jm}\frac{N_j}{N} {\widehat W}(g;jmm)&=&|g><g|;\label{313c}\\
{\rm Tr}[{\widehat W}(g;jmm^\prime)]&=& \delta_{mm^\prime};\label{313d}\\
{\rm Tr}[{\widehat W}(g;jmn)~{\widehat W}(g^\prime;j^\prime m^\prime n^\prime)
^\dagger]&=& 
\frac{N}{N_j}\delta_{g,g^\prime}\delta_{jj^\prime}
\delta_{mm^\prime}\delta_{nn^\prime};
\label{313e}
\eea
\end{subequations}
which are easily obtained by eqs.$(\ref{34a},\ref{27},\ref{26d})$ as 
appropriate.

We noted earlier that the point $(e;000)$ in phase space may be regarded as a
distinguished `origin'. The corresponding phase point operator 
${\widehat W}(e;000)$, from
eq.$(\ref{311})$, has the simple form and action 
\bea
{\widehat W}(e;000)&=&\sum_{g^\prime}|g^\prime><g^{\prime-1}|,\nonumber\\
{\widehat W}(e;000)|g>&=& |g^{-1}>.
\label{314}
\eea
We may call this as the inversion operation. This is the analogue, in the
present finite dimensional case treated algebraically, of the known result in
the case of continuous variables that the phase point operator at the origin
of phase space is a multiple of the parity operator.
\vskip2mm
\noindent
{\underline {\rm The Odd Prime Case}}
\vskip2mm

In case $N$ is an odd prime, the only choice for $G$ is to be an abelian
cyclic group of order $N$, so we exhibit it as 
\bea
G&=&\{e,g,g^2,\cdots g^{N-1}|g^N=e\}\nonumber\\
 &=&\{g^n|n=0,1,\cdots,N-1\}.
\label{315}
\eea
The UIR's are all one dimensional labelled by an integer $j=0,1,2,\cdots,N-1$:
\bea
D^j(g)&=&e^{2\pi ij/N}, \nonumber\\
D^j(g^n)&=&e^{2\pi ijn/N}.
\label{316}
\eea
The unique square root of $g^n$ is easily computed:
\bea
\sqrt{g^n}&=& g^{(n+N)/2},~~~ n=1,3,5,\cdots,N-2;\nonumber\\
          &=& g^{n/2},~~~~~~~~n=2,4,6,\cdots,N-1~.
\label{317}
\eea
The two Wigner distribution definitions $(\ref{35},\ref{39})$ coincide:
\be
W_{\psi}(g^n;j)=\sum_{n^\prime=0}^{N-1} \psi(g^{n-n^\prime}) e^{2\pi ij.
2n^\prime/N}\psi(g^{n+n^\prime})^*,
\label{318}
\ee
reproducing a known result \cite{5}. There is no `coarse graining' involved in 
the recovery $(\ref{37c})$ of the marginal momentum space probability 
distribution.

For general odd $N$, we may have again a unique $G$ in the abelian case, or
more than one distinct abelian choice. For example, for $N=9$, the two choices
$G=C_9$ and $G^\prime=C_3\times C_3$, where $C_n$ is the cyclic group of order
$n$, are nonisomorphic, so they lead to distinct definitions of the Wigner
distribution. On the other hand for $N=15$, $C_{15}$ and $C_3\times C_5$ are
isomorphic. The general pattern can be seen along these lines. 

The first case of a nonabelian $G$ occurs when $N=21$, some others occur for
instance when $N=27,39,45,55,57,63,75,\cdots$. It is in these cases that the
`coarse graining' of the momentum space probability distribution seen in
eqs.$(\ref{37c},\ref{310c})$ in two different ways, is involved. 
We examine this in some detail in the next section.

\section{Analysis of the Nonabelian Case}

Let $N$ be an odd nonprime such that a nonabelian group $G$ of order $N$
exists. In that case some further interesting properties of the two
definitions of Wigner distributions given in Section 3 can be exhibited. To
begin with, by a simple change of variable $g^\prime\rightarrow 
g^{-1}g^{\prime}g$ in equation eq.$(\ref{39})$ we can put
the second definition into the form
\be
W_{\psi}^{~\prime} (g;jnn^\prime) =\sum_{g^\prime} \psi(g^{\prime-1}g) 
D^{j}_{n n^\prime}(g^{-1}g^{\prime~ 2}g)\psi(g^\prime g)^*.
\label{41}
\ee

Comparing with $W_{\psi} (g;jmm^\prime)$ in eq.$(\ref{35})$, and for 
each $g$ and $j$ treating the two Wigner distributions as $N_j \times N_j$ 
matrices, we see that 
\be
W_{\psi}^{~\prime} (g;j)=D^{j}(g^{-1})W_{\psi}(g;j)D^{j}(g),
\label{42}
\ee
making clear the relation between them in detail.

Next we turn to the coarse graining of the momentum space probability
distribution involved in eqs.$(\ref{37c},\ref{310c})$. It is in fact possible 
to define an
overcomplete or redundant Wigner distribution in the following manner:
\bea
w_{\psi}(g;jmn^\prime~ m^\prime n) &=&\sum_{g^\prime} \psi(g^{\prime-1}g) 
D^{j}_{m n^\prime}(g^{\prime}g) 
D^{j}_{n m^\prime}(g^{-1}g^{\prime}) \psi(g^\prime g)^* \nonumber\\
&=&\sum_{g^\prime} \psi(gg^{\prime-1}) 
D^{j}_{m n^\prime}(gg^{\prime}) 
D^{j}_{n m^\prime}(g^{\prime}g^{-1}) \psi(gg^\prime)^*.
\label{43}
\eea
From here the two actual Wigner distributions emerge by partial tracing:
\bea
W_{\psi}(g;jmm^\prime)&=&\sum_{n}w_{\psi}(g;jmn~ m^\prime n),\nonumber\\
W_{\psi}^{~\prime}(g;jnn^\prime)&=&\sum_{m}w_{\psi}(g;jmn^\prime~ m n).
\label{44}
\eea

The quantities $w_{\psi}(g;jmn^\prime~ m^\prime n)$ have the properties 
\bea
w_{\psi}(g;jmn^\prime~ m^\prime n)^*
&=&w_{\psi}(g;jm^\prime n~ mn^\prime);\nonumber\\
\frac{N_j}{N}\sum_{g}w_{\psi}(g;jmn^\prime~ m^\prime n)&=&
\psi_{jm^\prime n}\psi_{jmn^\prime}^{*};\nonumber\\
\frac{1}{N}\sum_{jmn}N_j w_{\psi}(g;jmn~ m n)&=&
|\psi(g)|^2.
\label{45}
\eea

These results are consistent with eqs.$(\ref{44},\ref{37},\ref{310})$, 
and the behaviours under left and
right group actions can be easily worked out. We now see that all the momentum
space probabilities (in a polarised form), without any coarse graining, can be
recovered from the distribution $w_{\psi}(g;jmn^\prime~ m^\prime n)$. 
However this distribution is
overcomplete, as is demonstrated by showing explicitly that it can be
reconstructed from, say, $W_{\psi}(g;jmm^\prime)$. Combining the 
definition $(\ref{43})$ with eqs.$(\ref{311},\ref{312})$ we
have:
\bea
w_{\psi}(g;jmn^\prime~ m^\prime n) &=&\sum_{g^\prime}
<\psi|g^\prime g><g^{\prime-1} g|\psi>
 D^{j}_{m n^\prime}(g^{\prime}g)
 D^{j}_{n m^\prime}(g^{-1}g^{\prime}) \nonumber\\
&=&\sum_{g^\prime}\sum_{j^{\prime\prime}m^{\prime\prime}n^{\prime\prime}}
\frac{N_{j^{\prime\prime}}}{N} D_{m^{\prime\prime}n^{\prime\prime}}^
{j^{\prime\prime}}(g^{\prime~2})^* D_{mn^\prime}^{j}(g^\prime g)
D_{nm^\prime}^{j}
(g^{-1}g^\prime)\times\nonumber\\&&~~~~~~~~~~~~~~~~~~~~~~~~~~
W_{\psi}(g;j^{\prime\prime}m^{\prime\prime}n^{\prime\prime}).
 \label{46}
\eea
Again it is easily checked that this is consistent with eqs.
$(\ref{42},\ref{44})$. This
clarifies the situation in the nonabelian case. The full momentum space
probability distribution $|\psi_{jmn}|^2$, more generally 
the polarised expressions
$\psi_{jm^\prime n}\psi_{jmn^\prime}^*$, are all obtainable 
in principle from the Wigner distribution $W_{\psi}(g;jmm^\prime)$, even
though at first sight eq.$(\ref{37c})$ seems to suggest that only the 
coarse grained
expressions $\sum_{n}\psi_{jm^\prime n}\psi_{jmn}^* $ can be recovered. 
It is indeed the case that $W_{\psi}(g;jmm^\prime)$ describes
$|\psi><\psi|$ completely. 

\section{Conclusions}
In this work we have proposed a way of associating a  phase space with 
an $N$-level  quantum system by taking the corresponding 
`classical' or comparison configuration space as consisting of the 
$N$ elements of a group $G$ of order $N$. A gratifying feature of our 
construction that emerges is that the phase space always consists of 
an $N\times N$ array independently of the nature of the group $G$. 
We have  explored the possibilities for defining Wigner distributions on 
such phase spaces and have shown that this turns out to be possible only 
when $N$ 
is odd. Further, we have shown that for such dimensions,  there exist 
two equally good choices for the Wigner distribution and have discussed 
the relation between them in detail. We have also shown that the two Wigner
descriptions can be put together into an extended Wigner distribution from
which they can be recovered by partial tracing. Such a synthesis, however, 
can be achieved only at the expense of introducing overcompleteness into 
the description. 

The most curious feature that emerges from  this work is the importance of
being odd. We hope to return to this and related questions elsewhere.

\end{document}